\documentstyle[epsfig]{europhys}

\def\And{{\rm and\ }}

\newif\ifboo \boofalse

\def\Review#1{\boofalse{\it #1},}
\def\Name#1{{\sc #1},}
\def\Vol#1{\ifboo Vol. {\bf #1}\else{\bf #1}\fi}
\def\Year#1{\ifboo #1\else(#1)\fi}
\def\Book#1{\bootrue{\it #1},}
\def\Page#1{\ifboo {\rm p. #1}\else{\rm #1}\fi}

\begin{document}
\euro{}{}{}{}
\Date{}

\shorttitle{Structural features of the ripple phase}
\title{Novel structural features of the ripple phase of phospholipids}
\author{Kheya Sengupta\inst{1}, V.A. Raghunathan\inst{1} and John Katsaras\inst{2}}  
\institute{ 
\inst{1} Raman Research Institute, Bangalore - 560 080, India.\\
\inst{2} National Research Council, Steacie Institute of Molecular Sciences,
Chalk River Laboratories, Chalk River,  
Ontario  
K0J 1J0, Canada.}

\rec{}{}
\pacs{
\Pacs{61}{10.Eq}{X-ray scattering}
\Pacs{61}{30.Eb}{ Experimental determinations of smectic, nematic, cholesteric, and other structures
}
      }

\maketitle

\begin{abstract}
We have calculated the electron density maps of the ripple phase of 
dimyristoylphosphatidylcholine (DMPC) 
and palmitoyl-oleoyl phosphatidylcholine (POPC) multibilayers at different
temperatures
and fixed relative humidity. 
Our analysis establishes, for the first time, the existence of an 
average tilt of the hydrocarbon
chains of the lipid
molecules along the direction of the ripple wave
vector, which we believe is responsible 
for the occurrence of asymmetric
ripples in these systems.
\end{abstract}
Lipids self-assemble in water to form a variety of lamellar phases
~\cite{tar,smith}.
The ripple or {\sl P$_{\beta^{\prime}}$} phase characterized by a
one-dimensional height modulation of the bilayers is seen in some
phospholipids under high hydration ~\cite{tar,jan}.
In the phase diagrams of these systems, it is sandwiched between 
the high-temperature {\sl L{$_\alpha$}} phase  and the low-temperature
{\sl L{$_{\beta^{\prime}}$}} phase ~\cite{smith,jan}. In the {\sl
L{$_\alpha$}} phase, the
hydrocarbon chains of the lipid molecules are in a molten state and the
in-plane ordering is liquid-like. On the other hand, in the {\sl
L{$_{\beta^{\prime}}$}} phase, the chains are predominantly in the 
all-{\it trans} conformation and are tilted with respect to the layer
normal.  
The chains are also ordered in the plane of the bilayer, but the 
exact nature of the degree of this ordering is yet to be
determined~\cite{tar,smith}.\\

Experimental studies on the ripple phase 
have established many of its structural features.  
Almost all x-ray studies show asymmetric ripples corresponding to an
oblique unit cell of the rippled bilayers~\cite{tar,jan,raghu_jon,ww}; 
though there have been some
reports of symmetric ripples corresponding to a rectangular unit cell 
~\cite{hen,yao}.
The latter have been shown to be metastable structures in some systems
~\cite{yao}.
It is well established that only
lipids that have a {\sl L{$_{\beta^{\prime}}$}} phase at lower
temperatures exhibit the ripple phase, indicating the importance of the
tilt of the chains in the formation of the ripples ~\cite{kir}. 
Determination  of the chain tilt in the ripple phase is of utmost 
relevance
as it is the key structural feature hitherto unknown.
But it has not been possible to obtain it directly from x-ray
diffraction patterns as the chain
reflections are rather diffuse, probably due to the presence of disordered
chains, ie., chains that are not entirely in the {\it trans}
conformation. NMR
and diffusion
experiments also detect a population of
disordered chains in this phase ~\cite{wit,sch}. 
Hence detailed information about chain tilt has to be deduced
from the electron density map of the system, calculated from x-ray
diffraction data.\\

The lack of knowledge of its structural features has
hindered the formulation of a satisfactory theory of the ripple phase;
none of the current theories is consistent with all
the experimental observations.
The most striking
disagreement concerns the occurrence of asymmetric ripples.
It has been proposed that molecular chirality is responsible for
such ripples~\cite{lub1}, but experiments indicate otherwise
~\cite{raghu_jon,sen1}.\\

Recently, Sun et al.~\cite{nagle} have calculated the electron density
map of the ripple phase of DMPC using the x-ray data of Wack and Webb
~\cite{ww}. 
They find
that the ripples  have a saw-tooth shape, with the bilayer thickness
in the minor arm being much smaller than that in the major arm. 
Further, the electron
density in the headgroup region along the major arm is much higher than
that along the minor arm. This led them to  hypothesize that the chain
organization in the major arm is like in the {\sl L{$_{\beta^{\prime}}$}} 
phase and that in the minor arm is like in the  {\sl L{$_\alpha$}}
phase. 
However, this conjecture is not supported by other experiments.
For example, self-diffusion in the
ripple phase is found to be highly anisotropic, with a fast component that
is 4-5
orders of magnitude faster than the slow component ~\cite{sch}; but the
fast component
itself is about 2-3 orders of magnitude smaller than that in the {\sl
L{$_\alpha$}} phase. Thus the authors of Ref. 10 conclude that
although
the intramolecular hydrocarbon chain disorder may be substantial in the
fast bands, the intermolecular order in this region is not like that in
the {\sl L{$_\alpha$}} phase. \\

In view of this discrepancy, we have calculated the electron density maps
of the ripple phase of DMPC and POPC. In addition to the x-ray data of
Ref. 5, we have used data from oriented samples at different
temperatures and fixed relative humidity.
We find that the ripples in both these systems have a saw-tooth shape,
with the ratio of the lengths of the two arms essentially independent of
temperature.
If the molecules in the short arm were in the {\sl L{$_\alpha$}} phase,
the length of this arm would be expected to increase as the {\sl L{$_\alpha$}}
phase is approached from below. This is contrary to what we 
see. Further, the 
difference in the bilayer thickness and the electron density in the
two arms can be largely accounted for in terms of an average tilt of the
chains along the direction of rippling, which we believe is
responsible  for the occurrence of asymmetric ripples in these systems.
These results are clearly important  
for a  satisfactory theoretical description of this
phase.\\

We have adopted the modeling and least squares fitting 
procedure developed by Sun et al.~\cite{nagle} to calculate the electron
density maps. The unit cell parameters of the two-dimensional
oblique lattice are the two vectors {\bf a} and {\bf b}, and the
angle $\gamma$. In terms of the ripple wave length $\lambda$
and the lamellar spacing d, the two lattice vectors can be expressed as:
{\bf a} = d cot $\gamma$ \^{x} + d \^{z}, and {\bf b} = $\lambda$ \^{x}.
Here \^{x} is the direction of the ripple
wave vector and \^{z} is the direction of the average layer
normal (see Fig.1). $\lambda$, d, and $\gamma$ are directly measured from
the
diffraction pattern.
The electron density within the unit cell, $\rho$(x,z), is described as
the
convolution of a ripple contour function C(x,z) and the transbilayer
electron density profile T$_\psi$(x,z). 
C(x,z) = $\delta$(z-u(x)), where u(x) describes the ripple profile and 
is taken to
have the form of a saw-tooth with  
peak-to-peak amplitude A. $\lambda_{1}$ is the
projection of the longer arm of the saw-tooth on the x-axis. T$_\psi$(x,z) 
gives
the electron density at any point (x,z)  along a straight line,
which
makes an angle $\psi$ with the z-axis. The electron density in the methylene
region of the bilayer is close to that of water and is taken as zero.\\
 
We have used three models for 
T$_\psi$(x,z), 
two of which are equivalent to the SDF and M1G models of
Ref. 13. In model I, it is taken as consisting of two delta
functions
with positive coefficients $\rho_{H}$,
corresponding to the headgroup regions separated by a
distance L, and a central delta function with negative
coefficient of magnitude $\rho_{M}$,
corresponding to the methyl region. The six adjustable parameters in 
model I are: A, $\lambda_{1}$, $\psi$, $\rho_{H}$/$\rho_{M}$,
L and a normalizing factor.
In model II, the delta
functions representing the head and methyl groups are replaced
with Gaussians of width $\sigma_h$ and $\sigma_m$, respectively.
The electron density in the minor arm is allowed to be different by a
factor $f_1$ from that in the major arm. The region where the two arms
meet is modeled as a wall with an electron density differing by a factor
$f_2$ from the rest of the arm. The wall thickness in this model is
fixed at a small value. Thus there are 10
adjustable parameters in this model.
It is using these two models that Sun et al. ~\cite{nagle} find that
the
bilayer
thickness along the local layer normal is different in the two arms of the
ripple. But this result
is built into these models as the parameter L, which is the thickness of
the bilayer along a direction that makes an angle $\psi$ with
the z-axis, is taken to be the same in the two arms.
Therefore, in model III, we remove this constraint and allow L as well as
$\sigma_h$, $\sigma_m$ and $\rho_{H}$/$\rho_{M}$
to be different in the two arms of the ripple.
Further, the wall between the two arms is taken to have a variable width 
w. This model has 15 adjustable parameters.
Minimization is done by iterative least squares fitting
with respect to six variables at a time ~\cite{num}. \\

The structure factors at the observed (h,k) values are calculated
using each of the above models. These are then
compared with the observed structure factors and
a chi-square value is obtained, which is subsequently minimized by varying
the adjustable parameters in the model. The phase of each of the Bragg
reflections is obtained from the structure factors calculated from the
converged model. These calculated phases are combined with the observed
magnitudes of the structure factors and inverse Fourier transformed
to get the electron density map of the system.            
We have used the x-ray diffraction data of Wack and Webb~\cite{ww} 
from powder
samples of DMPC as well as our
data from oriented films of {\it l}-DMPC, {\it dl}-DMPC and POPC.
Details of the experimental procedure are discussed elsewhere
~\cite{raghu_jon}. Relevant
geometric intensity corrections were applied to the data from oriented
samples, but
absorption
corrections could not be applied as the sample thicknesses are not
accurately known. We have confirmed by assuming reasonable values
of the sample thickness, that these corrections do not significantly
change the calculated electron density profiles. However, in the absence
of these corrections, 
we are unable to analyze these data using models II
and III, due to the lack of convergence resulting from the
large number of adjustable parameters in these models.\\

The converged values of the different parameters in the three models are
given
in table 1, along with the crystallographic R factor. The precision
of these parameters is about 0.1 \%. The data from Ref.
5 were used in these analyses.
In accordance with the results of Ref. 13, 
we find that  all the phases, except those of the three relatively
weak (0,k) reflections,
obtained using model II are the same as those obtained from model I.
These phases are also the same as those obtained from the SDF and 
M1G models in Ref. 13.
Model III gives only a marginally better fit to the data,
the chi-square being only about 2\% lower. 
The phases of all the reflections other than the (0,k) reflections are
the same as those found from models I and II. For the (0,k)
reflections this
model gives a combination of phases that is different from the ones
obtained from
models I and II.
The values of L in the two arms are almost
the same and equal to that obtained from model II.
This is also true of the other parameters which are allowed to be
different in the two arms. However, model III
gives a slightly higher value of 0.7 for $f_1$. 
As discussed below,
this factor can be accounted for in terms of the chain tilt, without
resorting to the assumption of a  {\sl L{$_\alpha$}}-like organization
in the minor arm. The low $\chi^2$ and R values for models II and III,
and the absence of any physically unacceptable features in the electron
density map (see Fig.2) indicate that these models closely represent
the true structure of the system.\\

The electron density map of the ripple phase of DMPC, calculated with the
data of Ref. 5 is shown in Fig.2. The ripples clearly have a
saw-tooth
shape, with an offset between the two leaves of the bilayer.
The simplest explanation for this offset is an average tilt of the
chains along
the rippling direction; such an offset cannot be expected if the
tilt were in a plane normal to the rippling direction. The tilt angle
$\psi$
is found to be approximately equal to (${\gamma}$-$\frac{\pi}{2}$).
Further confirmation of the existence  of an average tilt along this
direction comes from the
fact that the value of L
is almost equal in the two arms 
and is comparable 
to twice the length of a fully stretched DMPC molecule. 
If it is assumed that the chains are tilted at an angle
 $\psi$ with the z-axis, their tilt with respect
to the local layer normal can be calculated from the shape of the ripple.
Using the values of the structural parameters given in table I, the tilt
angle with respect to the local layer normal turns out to be 1.6$^o$ and
34.5$^o$ in the longer and shorter arms,
respectively. 
The tilt in the short arm is comparable to that 
found in the {\sl L$_{\beta^{\prime}}$} phase.
Since the area per molecule is inversely proportional to the cosine of
this angle, a value of 0.82 is obtained for $f_1$. 
This is in very good agreement with the value of 0.77 obtained from
the map for the ratio of the average electron densities in the headgroup
region of the longer and shorter arms. Thus an average
tilt of the chains along the rippling direction provides a
consistent explanation for many features of the electron density map.
This means that to a good approximation the height modulation of the
bilayers along the x-axis can be described as arising from a relative
sliding movement of
neighboring chains, with all the
chains lying the x-z plane and tilted by a constant angle $\psi$
with respect to the z-axis. The existence of an average chain tilt along
the rippling direction breaks the reflection symmetry of the bilayer in
the plane normal to it and
hence can be expected to be responsible for the asymmetric ripples
seen in this system. \\

We have also calculated the electron density maps of {\it l}-DMPC,
{\it dl}-DMPC and
POPC at different temperatures in the ripple phase, using data from
oriented films. The structural features of the ripples are found to be
similar to those obtained from the data of Ref. 5.
The maps of the chiral
and racemic DMPC samples were identical,
indicating the lack of
influence of
molecular chirality on the ripple structure ~\cite{sen1}.  The temperature
dependence of the
structural parameters of the ripples in DMPC are found to be very weak,
as in the case of dipalmitoyl
phosphatidylcholine (DPPC) ~\cite{ino}.
Contrary to what
is observed in freeze fracture experiments
~\cite{zasa_amp}, we find that 
the ripple shape has Fourier components higher than the second.        
Further,  we do not find a significant temperature dependence of the
amplitude of the ripples in contrast to what is reported in
Ref. 17.\\

The electron density map of the ripple phase of POPC is shown in Fig. 3.
The ripple shape is very similar to that of DMPC. It also has a
saw-tooth shape and an offset between the monolayers indicating an
average 
tilt in the direction of rippling.
In POPC, the angle ${\gamma}$ is much larger than in DMPC,
whereas the wave length and layer spacing are comparable.
Unlike those of DMPC, the structural
features of the ripple phase of POPC vary significantly with temperature,
as shown in table 2. In the absence of absorption corrections, the 
fits are not as good as in the case of DMPC data of Ref. 5.
As mentioned earlier,
we find that the electron density profiles are insensitive 
to these corrections. Hence the values of the last two parameters
quoted in the table are the ones estimated from the electron density maps. 
The layer spacing decreases slowly and $\gamma$ increases steadily
as temperature is increased. 
The ripple wavelength first decreases 
and then suddenly increases to a large value just below the transition.
These trends are very similar to those seen in DPPC ~\cite{ino}, but in
POPC the temperature dependence is very much pronounced.
The amplitude, except near the L$_{\alpha}$ transition,
is about half that of DMPC.
In both DMPC and POPC the ratio of the projected lengths of the
major and minor arms is about 2 and is
essentially insensitive to
temperature. This observation further supports the view that the
chain organization in the minor arm
is not like that in the {\sl L{$_\alpha$}} phase.\\

All the freeze fracture studies of the ripple phase show ripples
oriented over
micrometer-sized regions ~\cite{tar,ino,zasa_amp,lun,rup}. 
Since the chain tilt is locked to the rippling direction,
this implies long range order of the tilt
direction. These experiments also show ripples oriented only along three
directions, each at an angle of approximately 120$^o$ from the other two,
indicating 
a six-fold symmetry in the 
underlying bilayer structure.
Thus the in-plane ordering of the molecules in the bilayer is at least 
hexatic.\\

In conclusion, we have calculated the electron density maps of the ripple
phase of DMPC and POPC. The shape of the ripples in these two systems are
very similar, with both of them exhibiting asymmetric ripples. 
We have been able to
establish the
existence of an average chain tilt in the direction of rippling, which
is probably responsible for the asymmetric ripples seen in these
systems. 
\\

We thank Y. Hatwalne, K. Usha, and J. F. Nagle for discussions and HKL
Research Inc. for the use of their software.\\

\noindent

\begin{table}
\caption{The values of the structural parameters obtained from the three
models.}
\noindent
\begin{tabbing}
000\=  100000000000 \= 10000000000 \= 10000000000 \= 10000000000 \kill
\> Parameter  \> Model I \> Model II \> Model III \\ 
\> A($\AA$) \> 19.7 \> 18.7 \> 19.1 \\
\> $\lambda_1$($\AA$) \> 106.2 \> 101.0 \> 101.5 \\
\> $\psi$ \> 5.7$^o$ \> 9.9$^o$ \> 9.1$^o$ \\
\> L($\AA$) \> 41.0 \> 37.0 \> 37.2,37.0*\\
\> $\rho_h / \rho_m$  \> 1.1 \> 0.9 \> 1.1,1.1*\\
\>  $f_1$ \> - \> 0.6 \> 0.7\\
\> $f_2$ \> - \> 0.9 \> 0.65 \\
\>  $\sigma_h$ \> - \> 4.9 \> 4.6,4.7* \\
\>  $\sigma_m$ \> - \> 13.3 \> 9.4,10.0* \\ 
\> w($\AA$) \> - \> 2.0(fixed) \> 9.2 \\
\>  $\chi^2$ \> 692.5 \> 217.6 \> 213.6 \\
\>  R \> 0.172 \> 0.083 \> 0.087 
\end{tabbing}
\noindent
{\small * indicates the values in the shorter arm}\\
\label{Tab1}
\end{table}

\begin{table}
\caption{Temperature variation of the structural parameters of the ripple
phase of POPC at  75\%RH.} 
\begin{tabbing}
000\=  100000000 \= 10000000 \= 10000000 \= 10000000
\= 10000000 \= 10000000  \kill
\> T($^o$C) \> d($\AA$)  \> $\gamma ^o$ \> $\lambda (\AA)$ \> A ($\AA$) \>
$\frac{\lambda_1}{(\lambda-\lambda_1)}$ \\
\> 13.0 \> 58.3$\pm$0.1 \> 116$\pm$1 \> 200$\pm$2 \> 10$\pm$1 \>
2.0$\pm$0.1 \\
\> 13.5 \> 58.0 \> 119 \> 170 \> 9 \> 1.8 \\
\> 14.5 \> 57.3 \> 124 \> 143 \> 5 \> 2.6  \\
\> 15.0 \> 56.4 \> 133 \> 266 \> 5 \> 1.9
\end{tabbing}
\label{Tab2}
\end{table}

\begin{figure}
\hbox{\centerline{\psfig{figure=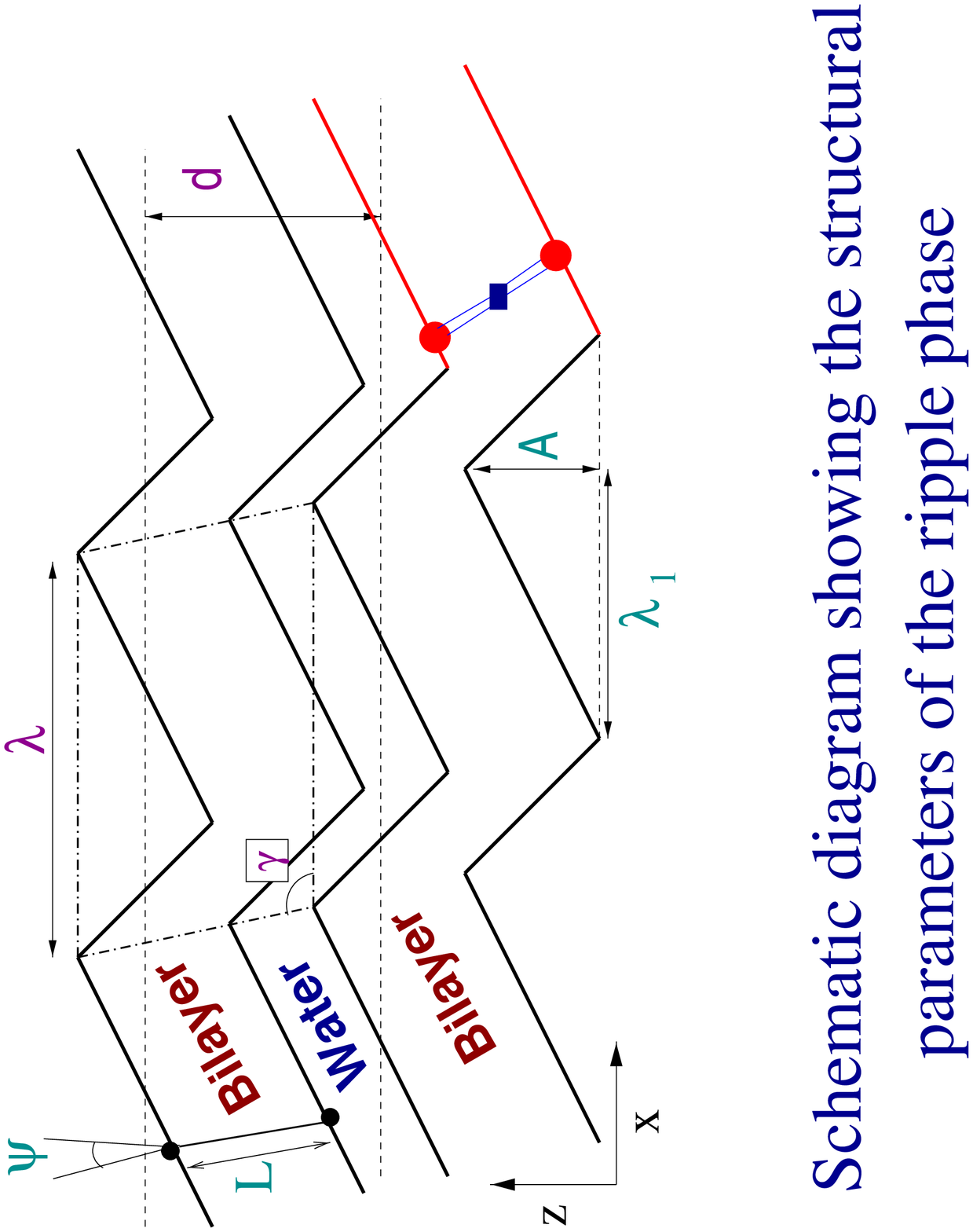,width=4.1 cm,angle=-90}}}
\caption{Schematic showing the structural parameters of the ripple
phase.}
\label{fig1}
\end{figure}
\begin{figure}
\hbox{\centerline{\psfig{figure=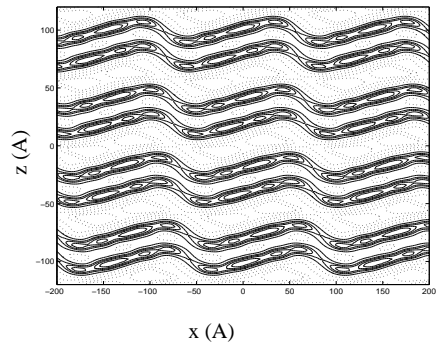,width=4.1 cm,angle=-90}}}
\caption{The electron density map of the ripple phase of DMPC obtained
using the data from Ref.13. T=18.2$^o$C and the volume fraction of water
is 0.263. $\lambda$=141.7{\AA} and $\gamma$=98.4$^o$. The positive
(negative) contours are represented by solid (dotted) lines.
The regions with positive electron density correspond to the head groups.
Note that the thickness of the bilayers is about 40 {\AA}, whereas that of the
water region is about 20 {\AA}.}
\label{fig2}
\end{figure}
\begin{figure}
\hbox{\centerline{\psfig{figure=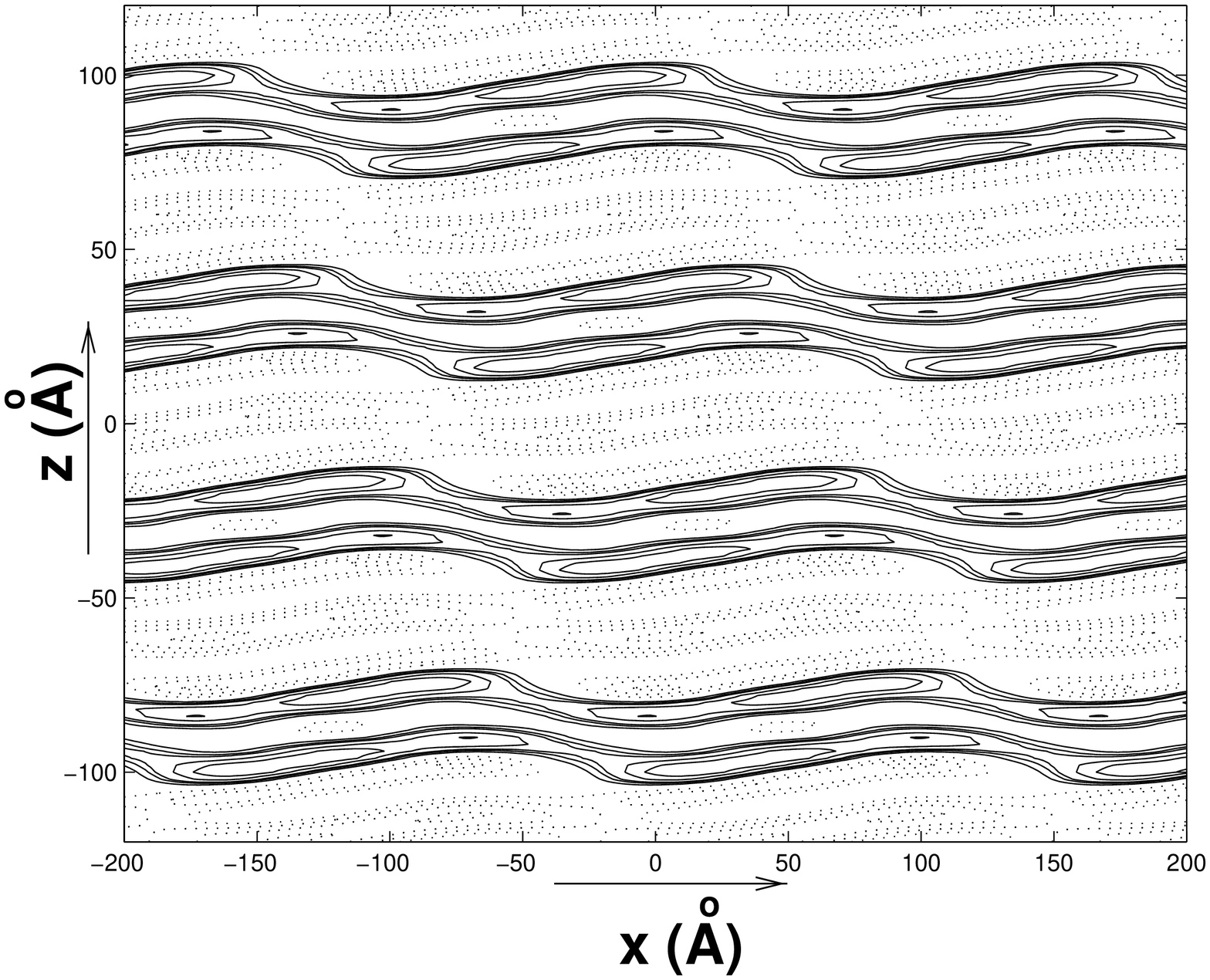,width=4.2 cm,angle=-90}}}
\caption{The electron density map of the ripple phase of POPC obtained
using x-ray data from oriented samples. T=13.5$^o$C and RH=75\%.}
\label{fig3}
\end{figure}


\vspace{-1.0 cm}

\begin{thebibliography}{99}
\bibitem{tar}
\Name{A. Tardieu, V. Luzzati \And F. C. Reman.}
\Review{J. Mol. Biol.}
\Vol{75}
\Year{1973}
\Page{711}


\bibitem{smith}
\Name{G.S. Smith, E.B. Sirota, C.R. Safinya \And N.A. Clark.}
\Review{Phys. Rev. Lett. }
\Vol{60}
\Year{1988}
\Page{813}

\bibitem{jan}
\Name{M. J. Janiak, D. M. Small \And G. G. Shipley.}
\Review{J. Biol. Chem.}
\Vol{254}
\Year{1979}
\Page{6068}

\bibitem{raghu_jon}
\Name{J. Katsaras \And V. A. Raghunathan.}
\Review{Phys. Rev. Lett.}
\Vol{74}
\Year{1995}
\Page{2022 }

\bibitem{ww}
\Name{D. C. Wack \And W. W. Webb.}
\Review{Phys. Rev. A }
\Vol{40}
\Year{1989}
\Page{2712}

\bibitem{hen}
\Name{M.P. Hentschel \And F. Rustichelli .}
\Review{Phys. Rev. Lett. }
\Vol{66}
\Year{1991}
\Page{903}

\bibitem{yao}
\Name{H. Yao, S. Matuoka, B. Tenchov \And I. Hatta .}
\Review{Biophys. J.}
\Vol{59}
\Year{1991}
\Page{252 }

\bibitem{kir}
\Name{S. Kirchner \And G. Cevc.}
\Review{Europhys. Lett. }
\Vol{28}
\Year{1994}
\Page{31}

\bibitem{wit}
\Name{R.J. Wittebort, C.F. Schmidt \And R.G. Griffin.}
\Review{Biochemistry}
\Vol{20}
\Year{1981}
\Page{4223}


\bibitem{sch}
\Name{M.B. Schneider, W.K. Chan \And W.W. Webb.}
\Review{Biophys. J. }
\Vol{43}
\Year{1983}
\Page{157}

\bibitem{lub1}
\Name{T. C. Lubensky \And F. C. MacKintosh.}
\Review{Phys. Rev. Lett. }
\Vol{71}
\Year{1993}
\Page{1565}
\Name{C.-M. Chen, T.C. Lubensky \And F.C. MacKintosh.}
\Review{Phys. Rev.
E }
\Vol{51}
\Year{1995}
\Page{504}

\bibitem{sen1}
\Name{K. Sengupta, V.A. Raghunathan \And J. Katsaras. }
\Review{Phys. Rev. E}
\Vol{59 (2)}
\Year{1999}
\Page{2455}


\bibitem{nagle}
\Name{W.-J. Sun, S. Tristram-Nagle, R. M. Suter \And J. F.
Nagle.}
\Review{Proc. Natl. Acad. Sci. USA}
\Vol{93}
\Year{1996}
\Page{7008}

\bibitem{num}
\Name{W.H. Press, S.A. Teukolsky, W.T.Vellerling \And
  B.P. Flannery.}
\Book{Numerical Recipes, Cambridge University Press, 1997} 

\bibitem{ino}
\Name{Y.Inoko, T. Mitsui, K. Ohki, T. Sekiya \And Y. Nozawa.}
\Review{Phys.
Stat. Sol. (a)}
\Vol{61}
\Year{1980}
\Page{115}

\bibitem{zasa_amp}
\Name{J. T. Woodward \And J. A. Zasadzinski.}
\Review{Phys. Rev.
E }
\Vol{53}
\Year{1996}
\Page{R3044}

\bibitem{lun}
\Name{E.J. Luna \And H.M. McConnell.} 
\Review{Biochim. Biophys. Acta .}
\Vol{466}
\Year{1977}
\Page{381}

\bibitem{rup}
\Name{D. Ruppel \And E. Sackmann .} 
\Review{J. Phys. (Paris). }
\Vol{44}
\Year{1983}
\Page{1025}
 
\end{thebibliography}
\end{document}